# RF transport

*Stefan Choroba*
DESY, Hamburg, Germany

**Abstract**
This paper deals with the techniques of transport of high-power radiofrequency (RF) power from a RF power source to the cavities of an accelerator. Since the theory of electromagnetic waves in waveguides and of waveguide components is very well explained in a number of excellent text books it will limit itself on special waveguide distributions and on a number of, although not complete list of, special problems which sometimes occur in RF power transportation systems.

## 1 Introduction

The task of a radiofrequency (RF) power transportation system in an accelerator is to transport the RF power generated by a RF power source to the cavity of an accelerator. Sometimes it is necessary to combine the power of several RF sources and very often it is necessary to transport the power to not just one cavity but a number of cavities. It is usually the aim to transport the power with high efficiency and high reliability.

Different types of transportation systems can be considered. Parallel wires or strip lines can be used to transport electromagnetic waves, but they cannot be used for high-power RF transport because of their low power capability due to radiation into the environment or electrical breakdown above a certain power level. Coaxial lines and hollow waveguides can be used to transport high-power RF. Coaxial lines are used at lower frequencies up to some 100 MHz and power of some 10 kW. Hollow waveguides typically are used for frequencies above some 100 MHz, power levels of more than some 10 kW up to several 10 MW and distances of several metres. Coaxial lines are used at these parameters only for short distances or when efficiency does not play the key role. This is due to losses in the inner conductor of the coaxial line, losses in the dielectric material or breakdown between the inner and outer conductor of the coaxial line at a high power level. There is of course no sharp line when to use coaxial lines or hollow waveguides. Although coaxial lines are in use at accelerators the next sections will concentrate on hollow waveguides since they are used in a larger number of applications.

## 2 Theory of electromagnetic waves in waveguides and of waveguide components

The theory of electromagnetic waves in waveguides and of waveguide components can be found in a number of excellent text books [1–4], school articles [5, 6] or school transparencies [7], of which some are listed in the references. Therefore, the theory will not be repeated here.

## 3 Waveguide distributions

Waveguides distributions are combinations of different waveguide components. They allow for power transport, combination and distribution of RF power. In addition they protect the RF power source from reflected power and allow for adjustment of RF parameters such as amplitude, phase or $Q_{ext}$.

Distributions can be complicated assemblies and sometimes different options exist to fulfil certain requirements.

The size of the waveguide is first determined by the RF frequency. It is then still possible to choose between two standard sizes. For the transport of RF power at, for example, 1.3 GHz one can consider WR650 or WR770 waveguides. The decision depends on considerations such as the maximum power to be transmitted, space availability for the installation, weight, cost or availability of waveguide components on the market. Whereas the maximum power demand might require larger size, space demands restrict us to smaller dimensions.

The type of distribution system depends on similar considerations. In addition, demands such as the required isolation between cavities (cross-talk), RF parameters to be controlled or adjusted and more requirements must be considered. It is therefore worthwhile to take into account all possible requirements as early as possible and to trade them off.

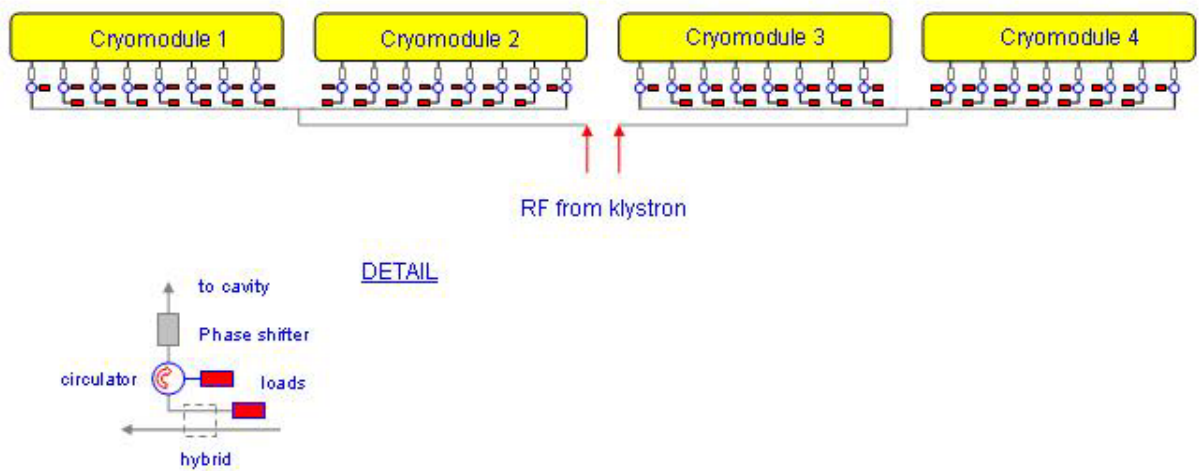

**Fig. 1:** Principle of a RF power transport system proposed for the TESLA linear collider

Figure 1 shows the principle of a waveguide distribution system which has been proposed for the TESLA linear collider. The RF power is generated by a 10 MW high-power klystron and extracted by two output windows towards four accelerator modules with eight superconducting cavities each. In order to achieve a gradient of 23.4 MV/m an input power of 231 kW per cavity is required. The total power produced by the RF power source must take into account losses in the waveguides and a regulation reserve. The power of each klystron waveguide arm is split again by a 3 dB hybrid. For each module a linear distribution system similar to that shown in Fig. 2 is used. Equal amounts of power are branched off for the individual cavities by hybrids with different coupling ratio. Isolators (three port circulators with loads) capable of 400 kW protect the power source from reflected power travelling back from the cavities during the filling time of the cavities or in the case of mismatch or breakdown in the cavities. Three stub tuners or phase shifters can be used for adjustment of phase and $Q_{ext}$.

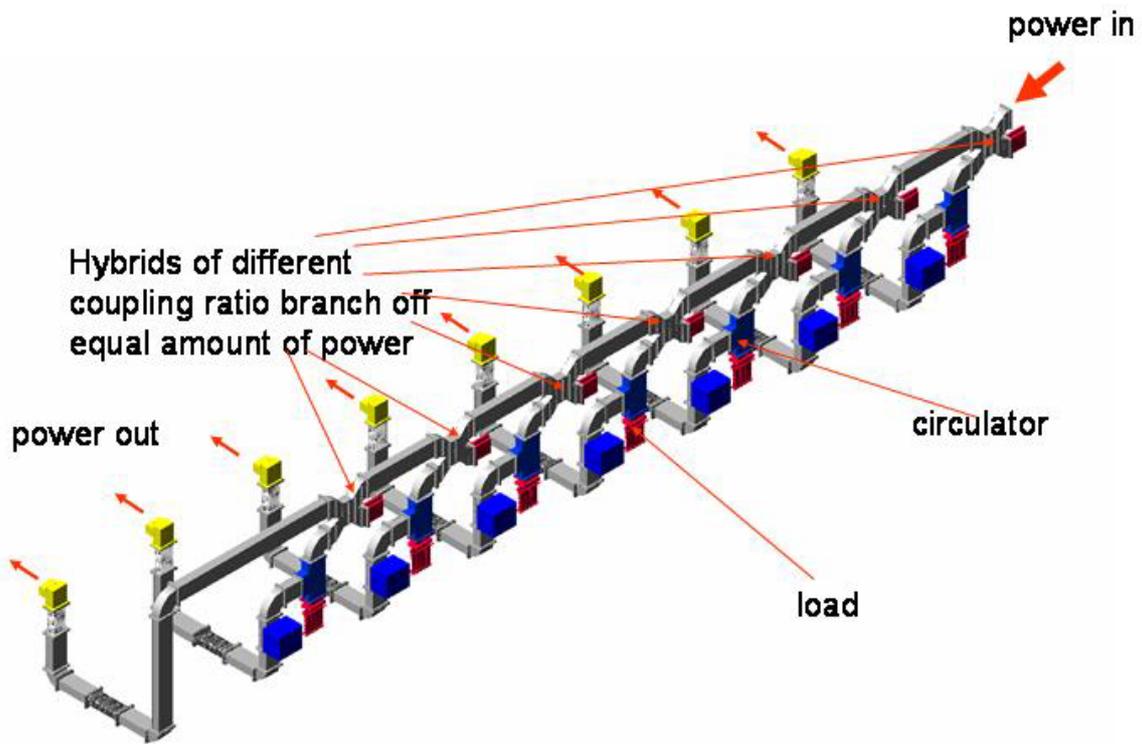

**Fig. 2:** Linear waveguide distribution system

Instead of a linear distribution system a tree-like system can be used (Fig. 3). The power is divided by shunt tees into several branches like the branches of a tree.

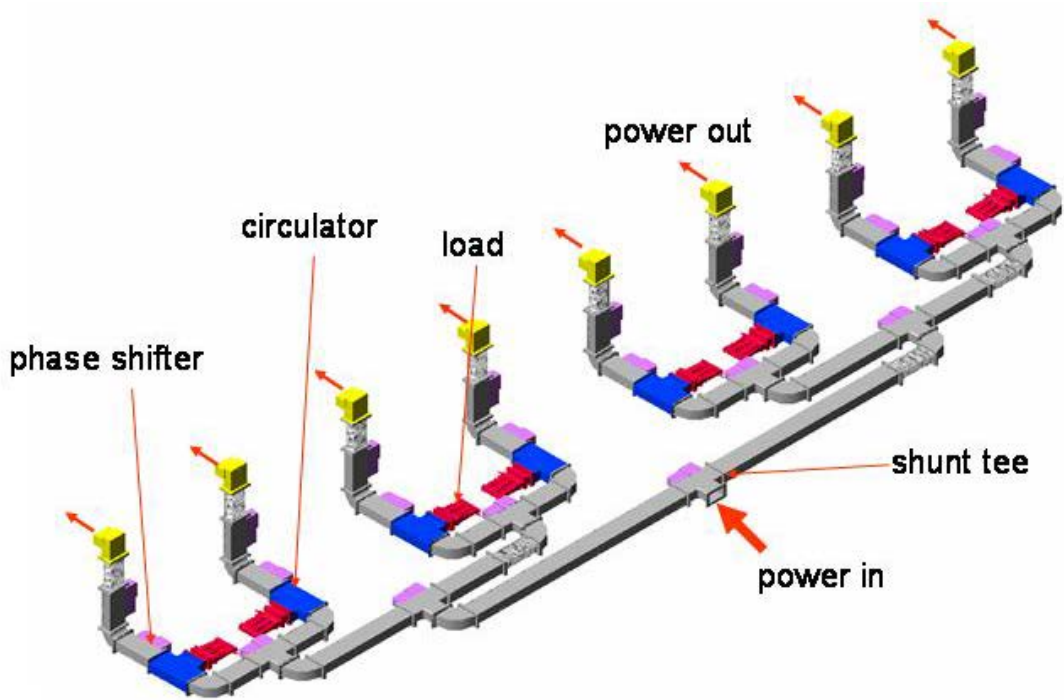

**Fig. 3:** Tree-like waveguide distribution system

In Fig. 4 examples of distributions meeting the requirements of power distributions for the FLASH facility at DESY are shown. The input power for the distribution can be up to 2.5 MW. Both distributions can be used to meet the requirements but because of certain advantages the first will be used in the future for the European XFEL. The latter and older system is a linear system and is used for the first accelerator modules at FLASH at DESY. The combined system proposed for the European XFEL and in use for the new RF distributions at FLASH makes use of asymmetric shunt tees of different coupling ratios. Equal amounts of power are branched off to a pair of cavities. By adjustment of tuning posts inside the tees the coupling ratios can be adjusted thus changing the branching ratio. The phase between a pair of cavities is pre-tuned by fixed phase shifters (straight waveguides with different waveguide width). The phase for the individual cavities can be adjusted by movable mechanical phase shifters after the symmetric shunt tee which splits the power for two cavities. Isolators in front of each cavity protect the power source from reflected power. This system has several advantages. Space and weight are reduced compared with the linear system. Phasing is much easier than in the purely linear system. Owing to the use of components of similar type the cost can be decreased. This system can also be pre-assembled and connected to an accelerator module before the module is installed in the accelerator tunnel, thus simplifying the entire installation procedure. An accelerator module with RF waveguide distribution can be seen in Fig. 5.

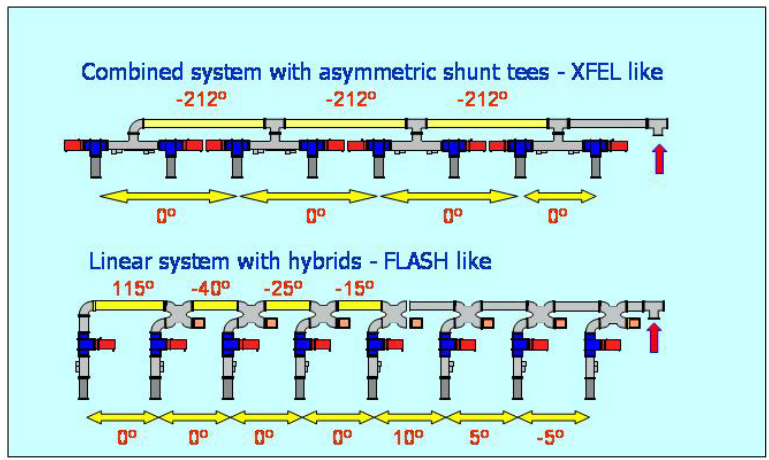

**Fig. 4:** Combined and linear RF waveguide distribution system

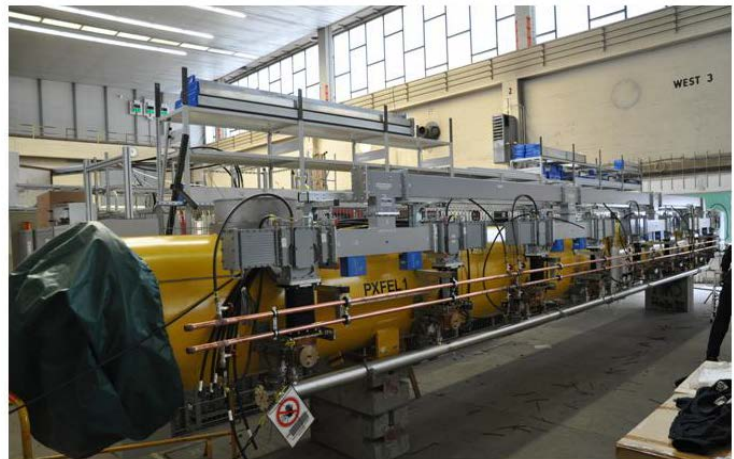

**Fig. 5:** Accelerator module and RF waveguide distribution

More information on the waveguide distribution systems for TESLA, FLASH and the European XFEL can be found in Refs. [8–12].

## 4    Limitations, problems and countermeasures

In this section some limitations, problems and possible countermeasures in RF power transportation systems are covered.

The power $P_{RF}$, which can be transmitted in a rectangular waveguide of size $a$ times $b$ in $TE_{10}$ mode of wavelength $\lambda$ is given by

$$P_{RF} = 6.63 \times 10^{-4} a[\text{cm}] b[\text{cm}] \left[ \lambda / \lambda_g \right] E[\text{V/cm}]^2$$

with

$$\lambda_g = \frac{\lambda}{\sqrt{1 - \left(\frac{\lambda}{2a}\right)^2}}$$

where $E$ (in V/cm) is the electrical field strength of the electromagnetic wave and $\lambda_g$ the guide wavelength. The maximum power is the power at the electrical breakdown limit $E_{max}$. In air $E_{max}$ is 30 kV/cm, which results in a RF power of 58 MW at 1.3 GHz in a WR650 waveguide. Experience shows that this power cannot be achieved. In practice, it is 5–10 times lower. The practical power limit is lower because of a variety of different reasons, for example the smaller size of the inner waveguide dimensions (e.g. within circulators), surface effects (roughness, steps at flanges, etc., see Fig. 6), dust in waveguides, humidity of the gas inside the waveguide, reflections (VSWR) or because of higher order modes (HOMs) in $TE_{nm}/TM_{nm}$.

These HOMs can be generated by the power source or by non-linear effects at high power in non-reciprocal devices such as circulators. If these modes are not damped, they can be excited resonantly and reach very high field strength above the breakdown limit. In order to damp HOMs, HOM dampers can be installed. These can be complicated and specially designed devices, which couple out and damp the HOMs only, leaving the fundamental mode, which is transmitted in $TE_{10}$, untouched. But sometimes a quick solution must be found. This can be accomplished by inserting small antennas at the small side of the waveguides (see Fig. 7). These antennas couple to a number of HOMs in $TE_{nm}/TM_{nm}$. Since the field of $TE_{10}$ is already small at the antenna position only a small amount of the power in the fundamental mode is coupled out. The antennas must be connected to loads, which must be able to handle the full amount of the power coupled out.

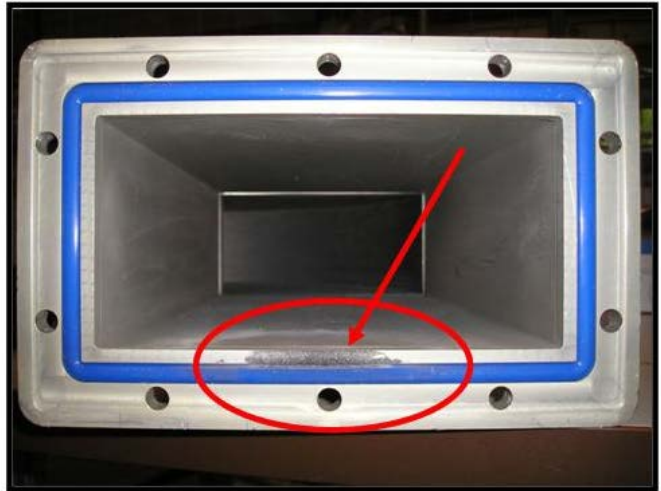

**Fig. 6:** Waveguide which has been damaged at the flange by breakdown

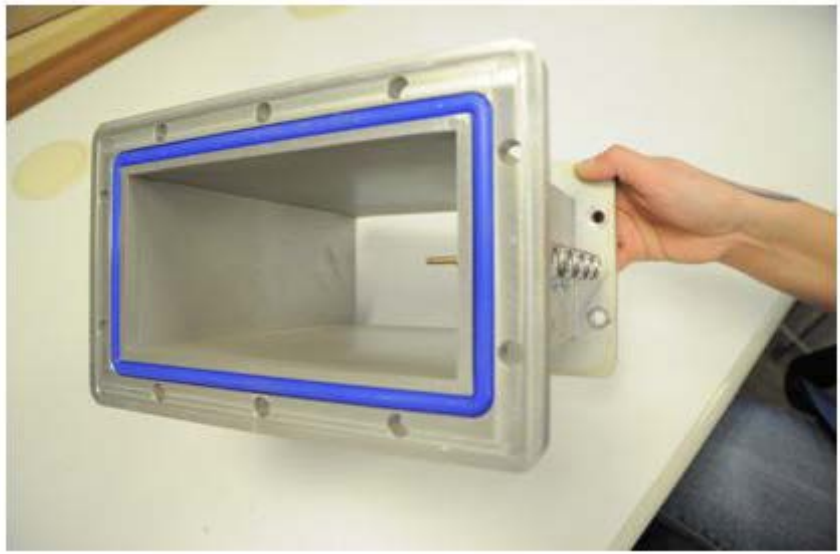

**Fig. 7:** Waveguide with damping antennas at the small side of the waveguide

One could increase the gas pressure inside the waveguide, which due to Paschen's law would increase the power capability, but this requires enforced and gas tight waveguides. In addition the pressure vessel rules applicable in many countries must be followed. By using $SF_6$ instead of air, which has $E_{max}$ = 89 kV/cm (at 1 bar, 20°C), the power capability can also be increased. The problem with $SF_6$ is that although it is chemically very stable it is a green house gas and if cracked in sparks it can, together with the hydrogen of the rest amount of water, form HF, which is a very aggressive acid. HF can produce fluorides of the metal of the waveguide walls which one can sometimes find as white powder in the waveguides (see Fig. 8). Other poisonous chemicals, e.g. $S_2F_{10}$, are also produced. Personnel maintaining waveguide components which have been operated in $SF_6$ have to observe a number of safety rules and wear personal protective equipment (see Fig. 9).

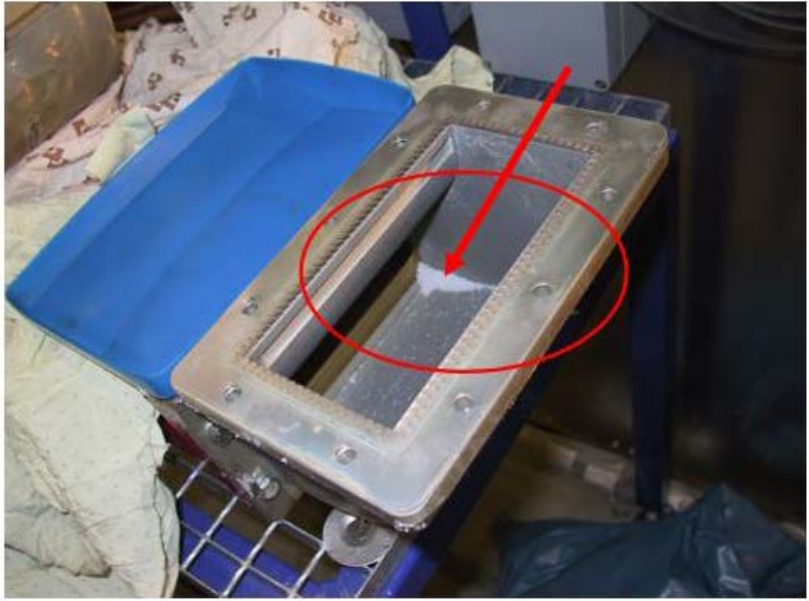

**Fig. 8:** Fluorides in a waveguide

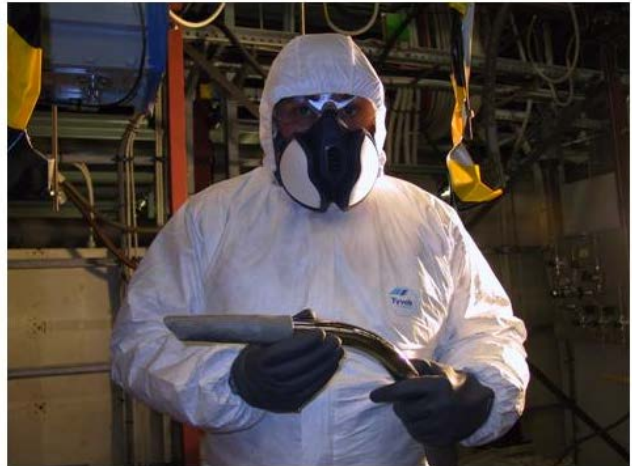

**Fig. 9:** Staff wearing protective clothing during work with $SF_6$ waveguides

## Acknowledgements

The author would like to thank two persons who among others supported him during the preparation of this lecture: Valery Katalev provided simulation results and pictures of the waveguide distributions; Ingo Sandvoss took many photographs.## References
[1] R.E. Collin, *Foundations for Microwave Engineering* (McGraw-Hill, New York, 1992).
[2] D.M. Pozar, *Microwave Engineering* (Wiley, New York, 2004).
[3] N. Marcuvitz, *Waveguide Handbook* (MIT Radiation Laboratory Series Vol. 10, McGraw-Hill, New York, 1951).